\documentclass[aps,prl,twocolumn,showpacs,a4paper]{revtex4}
\usepackage{graphicx}

\begin{document}





\title{ Superconductor--metal transition in an ultrasmall Josephson junction
biased by a noisy voltage source}  
\author{ E.B. Sonin}

\affiliation{ Racah Institute of Physics, Hebrew University of Jerusalem,
Jerusalem 91904, Israel} 

\date{\today} 

\begin{abstract}
Shot noise in a voltage source  changes the character of the quantum (dissipative)
phase transition in an ultrasmall Josephson junction: The {\em superconductor--insulator} transition transforms into the {\em
superconductor--metal} transition. In the metallic phase the $IV$ curve probes the
voltage distribution generated by shot noise, whereas in the superconducting phase it
probes the counting statistics of electrons traversing the noise junction.
\end{abstract}

\maketitle

The phenomenon of noise is continuing to be in the focus of modern mesoscopic
physics. In particular, a lot of attention is devoted to shot noise
\cite{Blanter}, which is related to discreteness of charge transport and
yields direct information on carrier charges.  Efforts of theorists were
invested to studying full counting statistics of shot noise, its
non-Gaussian character, and asymmetry (odd moments) 
\cite{Lez,Odd,BKN}. They are also objects of intensive experimental
investigations \cite{Reul,Rez}. This stimulated development of
effective methods of noise detection \cite{AK,SN,Delft,TN,exp,SN-T,Pek,grab,Scm,BF}. It
has been shown theoretically and experimentally that Coulomb blockade of an ultrasmall
Josephson junction is very sensitive to noise from an independent source
\cite{SN,exp,SN-T,Scm}. This can be used for noise spectroscopy. In the experiment they
studied the effect of shot noise on the low-voltage biased Josephson junction
\cite{SN,exp}. The source of shot noise   was a current through another junction parallel
to the Josephson junction. The current from the additional (noise) junction generated the
voltage drop on the shunt resistance, which was added to the voltage bias on the Josephson
junction. Originally this phenomenon was investigated al low noise currents,
when electron tunneling events produced a sequence of voltage pulses on the
shunt resistance well separated in time \cite{SN,exp,SN-T}. Later on the
theoretical analysis was expended on arbitrary high noise currents
\cite{Scm}. The analysis has shown that whatever high the current through
noise junction is, the voltage drop
produced by the noise current cannot be considered as an ideal voltage bias.
Thus the response of the Josephson junction to this ``noisy'' voltage source
is essentially different from that to the ideal  source of constant
voltage.

\begin{figure}
  \begin{center}
    \leavevmode
    \includegraphics[width=0.9\linewidth]{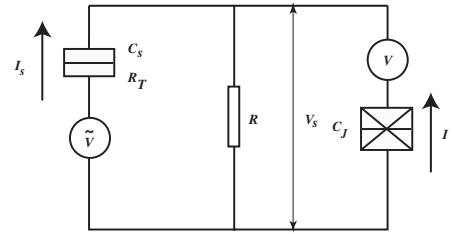}
    \caption{Electric circuit with two sources of constant voltages $V$ and $\tilde
{V}$. The Josephson junction with the capacitance $C_J$ is biased with the voltage $V$ and
the fluctuating voltage drop $V_s$ at the shunt. The latter originates from the
average current $I_s=\tilde V/R_T$ through the noise junction of capacitance $C_s$.}
  \label{fig1}
  \end{center}
  \end{figure}

This Letter investigates this interesting phenomenon. The analysis addresses
the general case when the Josephson junction is biased both with the ideal
and the noisy voltage source, but the focus is on the case, when  the
current through the additional  junction is the {\em only} source of the
voltage bias on the Josephson junction. The response of the Josephson
junction to this ``noisy'' voltage source is different from that
to the ideal (constant) voltage source in many aspects. However, the most
interesting outcome is that the well known ``superconductor--insulator''
transition \cite{Aver,SZ,IN}, which takes place at $\rho=R/R_Q=1$, transforms to the 
``superconductor--metal'' transition. Here $R_Q=h/4e^2=\pi
\hbar/2e^2$ is the quantum resistance for Cooper pairs. This means that at
$\rho>1$ the zero-bias conductance does not vanish, but remains finite as in a
junction between two normal metals.

Figure \ref{fig1} shows the electric circuit discussed in the paper. Two voltages can bias
the Josephson junction of capacitance $C_J$: (i) the constant voltage $V$ (ideal voltage
bias), and (ii) the fluctuating voltage drop $V_s$ at the shunt resistance $R$. The
average voltage $\bar V_s=I_s R$ is determined by the average current
$I_s$ through the additional noise normal junction of capacitance $C_s$ and resistance
$R_T$. It is assumed that $R_T \gg R$. Then $I_s =\tilde V/R_T$, the noise junction is
voltage biased and tunneling events at the junction are governed by the Poissonian
statistics. 

If only the ideal voltage bias is used
($I_s=0$), the $P(E)$ theory of incoherent tunneling of Cooper pairs yields the
following current through the Josephson junction
\cite{Aver,SZ,IN}:
\begin{equation} I={\pi eE_J^2 \over \hbar}[P_0(2eV)-P_0(-2eV)]~, 
 \label{IP}\end{equation} 
where the $P(E)$ function
\begin{equation} P_0(E)={1\over 2\pi \hbar} \int_{-\infty}^\infty dt
\exp\left[ J_0(t) +
\frac{iEt}{\hbar}\right]\, , 
    \label{P-E}    \end{equation} 
characterizes the probability to transfer
the energy $E>0$ to environment (or to absorb the energy $|E|$ from
environment if $E<0$). We restrict ourselves with the zero-temperature limit when the
phase-phase correlator, which determines the $P(E)$ function,  is given by
\cite{SN-T,Scm}
\begin{eqnarray} 
J_0(t) =\langle[ \varphi_0(t)
-\varphi_0(0)]\varphi_0(0)\rangle = \rho \left[-e^{t/\tau}\mbox{E}_1\left({t\over
\tau}\right)
\right. \nonumber \\ \left.
- e^{-t/\tau}\mbox{E}_1\left(-{t\over \tau}+i0\right)  -2 \ln{ t
\over \tau}
 -2\gamma -i\pi\right] \,,
     \label{J-0}    \end{eqnarray} 
where $\tau=RC$ is the relaxation time in the electric circuit, $C=C_J+C_s$,
$\gamma=0.577$ is the Euler constant, and
$\mbox{E}_1(z)=\int_1^\infty e^{-zt}dt/t$ is the exponential integral. The $P(E)$
theory is based on the time-dependent perturbation theory with respect to the small
Josephson coupling energy $E_J$ and uses the Golden Rule for calculation
of the tunneling probability. In addition, it is assumed that phase
fluctuations are Gaussian.  At $T=0$ $P(E)$ vanishes for
$E<0$ since it is the probability of the transfer of the energy $|E|$ from
the environment to the junction, which is impossible if $T=0$. The
subscript 0 points out that the phase fluctuations $\varphi_0$ and the
$P_0(E)$ function  are determined by the equilibrium Johnson-Nyquist noise. 

If the voltage bias is not ``ideal'', {\em i.e.}, the constant voltage
$V$ is supplemented with the fluctuating voltage drop $V_s$ at the shunt resistance
$R$, the $P_0(E)$ function in the expression for the current, Eq.(\ref{IP}), should be
replaced by a more general function \cite{SN-T,Scm}:
\begin{equation} 
P(2eV)={1\over 2 \pi \hbar} \int_{-\infty}^\infty
dt e^{J_0(t) }e^{i2eVt/\hbar}\left\langle\exp\left[  i\Delta
\varphi_s(t)\right]\right\rangle\, .
    \label{P-Vs}    \end{equation} 
The phase difference $\Delta\varphi_s(t)=\varphi_s(t)-\varphi_s(0)=(2e/\hbar)\int_0^t
V_s(t)$ is determined by the fluctuating voltage $V_s(t)$. Introducing the $P(E)$
function for shot noise,
\begin{equation} 
P_s(E)={1\over 2 \pi \hbar} \int_{-\infty}^\infty
dt e^{iEt/\hbar}\left\langle\exp\left[  i\Delta
\varphi_s(t)\right]\right\rangle\, ,
    \label{PEs}   \end{equation} 
the total $P(E)$ function is determined by a convolution of the two $P(E)$ functions
\cite{Heik}:
\begin{equation} 
P(2eV)=2 e \int_{-\infty}^\infty
dV_1P_0(2eV_1)P_s(2e[V-V_1])\, .
      \end{equation} 

The averaged phase correlator in the expression for the shot-noise
$P_s(E)$ function, Eq. (\ref{PEs}), is a value of the generating function
$\left\langle\exp\left[\xi\Delta\varphi_s(t)\right]\right\rangle$  at $\xi=i$. The
generating function determines all moments and cumulants of the random phase difference
$\Delta\varphi_s(t)$. For an ideal voltage bias of the noise junction this generating
function can be determined exactly \cite{Scm,HPS} keeping in mind that the electron
transport through the noise junction produces a sequence of random voltage pulses at
the shunt resistance $R$:
\begin{eqnarray} 
V_s(t)=\mbox{sign}(I_s)(e/C)\sum_i \Theta (t-t_i)
e^{-(t-t_i)/\tau}  ~,
   \label{V-ran}  \end{eqnarray} 
where $t_i$ are random moments of time when an electron crosses the junction. The
average time interval between tunneling events is $e/I_s$, and the number of
the events in a fixed time interval is governed by Poissonian statistics.
This sequence of voltage pulses generates the sequence of phase jumps: 
\begin{eqnarray}
\varphi_s(t)   =\mbox{sign}(I_s) \pi\rho \sum_i
\Theta (t-t_i)\left [ 1- e^{-(t-t_i)/\tau}\right]  ~.
     \end{eqnarray} 
The generating function for the random phase difference $\Delta
\varphi_s(t)$ is given by
\begin{equation}
\left\langle\exp\left[  \xi \Delta
\varphi_s(t)\right]\right\rangle=\exp\left[{I_s\tau\over e}\Phi(\xi
,t)\right]\, ,
    \label{phi-exp}  \end{equation} 
where
\begin{eqnarray}
 \Phi(\xi,t)=   - e^{\pi\rho \xi} \left[E_1(\pi\rho \xi) - E_1 \left(\pi\rho \xi 
e^{-t/\tau} \right)\right] -{t \over \tau}  -\gamma \nonumber \\ 
-\ln \left[-\pi\rho \xi\left(1- e^{-t/\tau} 
\right) \right]-E_1\left[-\pi\rho \xi\left(1- e^{-t/\tau} 
\right) \right]  ~.
      \label{phi}        \end{eqnarray}

In the high-impedance case $\rho\gg 1$,  the main contribution to the time integral in
Eq. (\ref{P-Vs}) for the $P(E)$ function comes from times $t \sim RC/\rho=R_QC$ much
shorter than $\tau=RC$ [see the analytic calculation of the ratchet current, Eq. (34), in
Ref. \onlinecite{Scm}]. Then the voltage does not vary essentially during the time
interval
$t$, i.e., 
$\Delta\varphi_s(t)\approx 2eV_s t/\hbar$, and the expression for
$\Phi(\xi,t)$ can be simplified:
\begin{eqnarray}
\Phi(\xi,t)=  -E_1 (-\pi\rho \xi t/\tau)  -\gamma-\ln (-\pi\rho \xi t/\tau)~.
       \label{shT}       \end{eqnarray} 
This means that the full statistics of the phase difference is identical to the
 full statistics of voltage fluctuations as found in Ref. \onlinecite{HPS}. Indeed, for
the sequence of random voltage pulses giving by Eq. (\ref{V-ran}),  the generating
function for voltage probability at the shunt is:
\begin{eqnarray} 
F(\nu) =\left\langle\exp\left[  \nu V_s C/e
\right]\right\rangle=\exp\left[{I_s \tau \over e}\Phi_v(\nu)\right]~,
         \label{GF-v}     \end{eqnarray} 
where 
\begin{eqnarray}
\Phi_v(\nu) =-E_1(-\nu) 
-\gamma -\ln(-\nu)~.
       \label{volt}       \end{eqnarray}
One can see that $\Phi(\xi,t)$ given by Eq. (\ref{shT}) is identical to $\Phi_v(\nu)$
in Eq. (\ref{volt}) with $\nu=\pi\rho \xi t/\tau$. Altogether this means that 
the voltage distribution generated by shot noise,
\begin{eqnarray}
p(V_s)={C\over 2\pi e}\int_{-\infty}^{\infty} dx e^{-ix
V_sC/e} F(ix)~,
      \label{volt-p} \end{eqnarray}
directly determines the shot-noise $P(E)$ function:
\begin{eqnarray}
P_s(E)={1\over 2e}p(-E/2e)~.
       \end{eqnarray}
Thus the total $P(E)$ function is the $P(E)$ function for the equilibrium noise
averaged over the voltage distribution generated by the noise current:
\begin{equation} 
P(2eV)= \int_{-\infty}^\infty
dV_1P_0(2eV_1)p(V_1-V)\, .
  \label{volt-dis}    \end{equation} 
This approach called ``the time-dependent $P(E)$ theory'' has already been used in
previous numerical simulations \cite{TDPE}. The present analysis has justified this approach for
the high-impedance environment $\rho\gg 1$, when the $P(E)$ function (current) probes
the voltage statistics. But the approach is not valid in the opposite case of the
low-impedance environment $\rho < 1$ (see below).

\begin{figure}
  \begin{center}
    \leavevmode
    \includegraphics[width=0.8\linewidth]{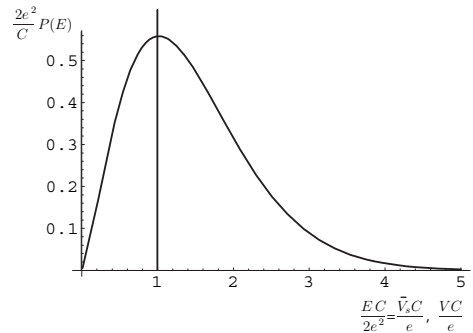}
    \caption{$P(E)$ function in the limit of high impedance $R\gg R_Q
$ for the ideal voltage bias ($E=2eV$, the
$\delta$-function peak at $V=e/C$) and for the noisy voltage bias
($E=2eV_s$).}
  \label{fig2}
  \end{center}
  \end{figure}

In the high-impedance limit $\rho =R/R_Q\to \infty$ the equilibrium $P(E)$ function can be
approximated with the $\delta$-function peak: $P_0(2eV)=\delta(2e	V- 2e^2/C)$ \cite{Naz}.
Then according to Eq. (\ref{volt-dis}) the $P(E)$ function (current) directly scans the
voltage probability distribution generated by shot noise and given by  Eq. (\ref{volt-p}):
\begin{equation} 
P(2eV)={1\over 2e}p\left({e\over C}-V\right)\, .
    \label{peak-appr}  \end{equation} 
The dependence of this function on $\bar V_s$ is plotted in Fig. \ref{fig2} for the
case when the Josephson junction is biased only with noisy voltage drop ($V=0$). In
contrast to the ideal voltage bias, which in the limit $\rho \to \infty$ yields the sharp
peak $P(2eV)=(1/2e)\delta(V-e/C)$, the noisy voltage bias yields the non-Gaussian broad
maximum. 

\begin{figure}
  \begin{center}
    \leavevmode
    \includegraphics[width=0.8\linewidth]{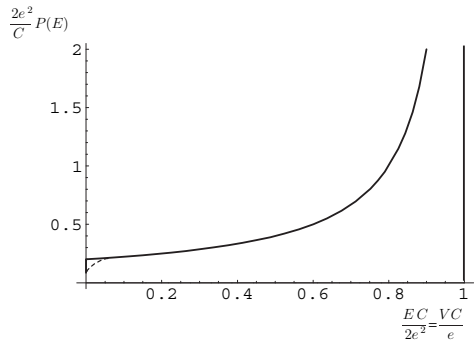}
    \caption{$P(E)$ function at weak noise current ($\bar V_sC/e=0.15$). The dashed
line shows smearing of the zero-bias jump with finite width of the peak in the
equilibrium $P_0(E)$ function.}
  \label{fig3}
  \end{center}
  \end{figure}

While the approximation based on Eq. (\ref{peak-appr}) is quite satisfactory for
calculation of the current dependence on noisy voltage bias $V_s$, it
is not sufficient for description of the effect of weak
shot noise ($\bar V_s \ll e/C$) on dependence on ideal voltage bias $V$, which
was studied in Refs. \onlinecite{SN-T,Scm}. In order to show it we linearize Eq.
(\ref{volt-p}) with respect to $\bar V_s$. Taking the integral by parts one obtain:
\begin{eqnarray}
p(V_s)={\bar V_s C^2\over 2\pi e^2}\int_{-i\infty}^{i\infty} d\nu e^{-\nu
V_sC/e} \Phi_v(\nu) \nonumber \\
={\bar V_s C\over 2\pi e V_s}\int_{-\infty}^{\infty}  e^{-ix
V_sC/e} {e^{ix}-1\over ix}dx \nonumber \\
={\bar V_s C\over 2\pi e V_s}\int_{-\infty}^{\infty}  
{\sin[x(1-V_sC/e)] +\sin(xV_sC/e)\over x}dx ~.
     \end{eqnarray}
This yields $p(V_s)=\bar V_s C/eV_s $ at $ 0<V_s<e/C$ and $p(V_s)=\bar V_s C/2eV_s $ if
$V_s=e/C$ exactly. Otherwise ($V_s<0$ or $V_s>e/C$) the voltage  probability vanishes. The
obtained probability density is the voltage distribution inside a single
voltage pulse $V_s(t)=(e/C)\exp (-t/\tau)$, and $p(V_s)\propto dt/dV_s$. Inserting
this voltage distribution into Eq. (\ref{peak-appr}) one obtains the $P(E)$ function
plotted as a function of the ideal voltage bias $V$ in Fig. \ref{fig3}. There are two
singularities on this dependence: a divergence near $V =e/C$ and a jump near the origin
$V=0$. But singularities are smeared out if one takes into account that the peak of
$P_0(E)$ at $E=e^2/C$ is not a $\delta$-function, but has a finite width $\sim
e^2/C\sqrt{\rho}$ (ignoring a logarithm factor). This more accurate approach, which was
used in Refs. \onlinecite{SN-T} and \onlinecite{Scm}, leads to smearing of the
zero-voltage jump onto voltages of the order of $e/C\sqrt{\rho}$ and to a finite linear
slope at $V=0$ (the dashed line in Fig. \ref{fig3}). The linear slope corresponds to
metallic conductance analytically calculated in Ref. \onlinecite{SN-T} for $\rho
\gg 1$ [see Eq. (20) there].

In the opposite limit of low-impedance environment  $\rho \ll 1$ the
most important contribution to the $P(E)$ function comes from long times $t \gg \tau$,
and the $P(E)$ function does not scan the voltage distribution anymore. In this limit we
need the asymptotic expression for the Johnson-Nyquist correlator, Eq. (\ref{J-0}):
\begin{eqnarray} 
J_0(t) =  -2\rho \left( \ln{ t\over \tau}+{i\pi\over 2}\right) \,,
          \end{eqnarray}
and for the logarithm of the generating function for the full phase statistics, Eq.
(\ref{phi}):  
\begin{eqnarray}
 \Phi(\xi,t)= {t\over \tau}(e^{\pi \rho \xi } -1) ~.
      \label{phi-Pous}     \end{eqnarray}
This expression corresponds to the Poissonian statistics of phase jumps, identical to
the Poissonian full counting statistics \cite{Scm,HPS}. Since $\rho \ll 1$ we can
expand in $\rho$. At the same time we can generalize this expansion on other possible
types of statistics. Then the phase correlator (the generating function at $\xi=i$)
can be written as 
\begin{equation}
\left\langle\exp\left[  i \Delta
\varphi_s(t)\right]\right\rangle=\exp\left[{i2e\bar V_st \over \hbar}\lambda\right]\,,
 \label{FS}   \end{equation}  
where
\begin{eqnarray}
 \lambda=\lambda_R+i\lambda_I\approx 1+{i\pi\rho \xi\over 2}{\langle\langle n^2
\rangle\rangle\over
\langle\langle n \rangle\rangle} -{\pi^2 \rho^2 \xi^2\over 6} {\langle\langle n^3
\rangle\rangle\over
\langle\langle n \rangle\rangle}~.
         \end{eqnarray}
Here $\langle\langle n^k \rangle\rangle$ are cumulants of the full counting
statistics,  $n$ being a number of electrons traversing the noise junction during the
time $t$. For the Poissonian statistics all cumulants are equal to the first cumulant 
$\langle\langle n \rangle\rangle$, which is the average number of electrons traversing
the junction. Inserting the expression (\ref{FS}) into  Eq. 
(\ref{P-Vs}) one obtains the $P(E)$ function in the low-impedance limit:
\begin{eqnarray} 
P(2eV)={2\over \pi\hbar}\int_0^\infty dt \left(\tau\over
t\right)^{2\rho}\sin(\pi \rho)
\nonumber \\ \times \mbox{Im}  \left\{ 
\exp\left[i\left({2eV t\over \hbar}+{2e\bar V_s t\over \hbar}\lambda \right)
\right]\right\}
\nonumber \\  \approx {2\tau \rho  \over \hbar}\left[\hbar\over 2e\tau(V+\bar V_s|\lambda|)\right]^{1-2\rho}\,.
    \label{Low}   \end{eqnarray} 
Though one cannot use this expression at very low voltages where the
perturbation theory in $E_J$ becomes invalid \cite{IG}, the ``superconductivity'' current
peak is present anyway, both for ideal and noisy voltage bias. Equation (\ref{Low})
demonstrates that in low impedance environment the $IV$ curve probes the counting
statistics even though the dependence  on high cumulants $k>2$ is not so pronounced.

In summary, shot noise in the voltage source dramatically changes the
character of the quantum (dissipative) phase transition in the ultrasmall
Josephson junction tuned by the environment impedance. For the ideal voltage
bias this is a transition from the superconducting state ($\rho=R/R_Q<1$) to the
insulator (Coulomb blockade, $\rho=R/R_Q>1$). In contrast, in the case of the noisy
voltage source  the transition at $\rho=R/R_Q=1$ is between the superconducting phase
and the metallic phase with finite zero-bias conductance.  This transition can be called
{\em superconductor--metal transition}. In the metallic phase the $IV$ curve is a
probe of the voltage distribution generated by shot noise, whereas in the
superconducting phase the $IV$ curve is probing the counting statistics for electrons
traversing the noise junction.

The author thanks Pertti Hakonen and Yuli Nazarov for interesting discussions.


\begin{thebibliography}{99}

\bibitem{Blanter} Ya.M. Blanter and M. B\"uttiker, Phys. Rep.
{\bf 336}, 1 (2000).
\bibitem{Lez} G.B. Lesovik, Pis'ma Zh.  Eksp. Teor. Fiz.  {\bf 60}, 806 (1994)[JETP Lett. {\bf 60},  820 (1994)];  L.S. Levitov, H. Lee, and G.B. Lesovik, J. Math. Phys. {\bf 37},
4845 (1996); L.S. Levitov and M. Reznikov, Phys. Rev. B {\bf 70}, 115305 (2004).
\bibitem{Odd} A. Shelankov and J. Rammer, Europhys. Lett. {\bf 63}, 485 (2003); D.B. Gutman and Y. Gefen, Phys. Rev. B {\bf 68}, 035302 (2003).
\bibitem{BKN}  C.W.J. Beenakker, M. Kindermann, and  Yu.V. Nazarov, Phys. Rev. Lett. {\bf 90}, 176802 (2003).
\bibitem{Reul} B. Reulet, J. Senzier, and D.E. Prober, Phys. Rev. Lett. {\bf 91},
196601 (2003); B. Reulet, cond-mat/0502077.
\bibitem{Rez} Yu. Bomze, G. Gershon, D. Shovkun, L.S. Levitov, and M. Reznikov, cond-mat/0504382.
\bibitem{AK} R. Aguado and L.P. Kouwenhoven, Phys. Rev. Lett. {\bf 84},
1986 (2000).
\bibitem{SN} J. Delahaye, R. Lindell, M.S. Sillanp\"a\"a, M.A. Paalanen,
E.B. Sonin, and P.J. Hakonen, cond-mat/0209076 (unpublished).
\bibitem{Delft} R. Deblock, E. Onak, L. Gurevich, and L.P. Kouwenhoven,
Science {\bf 301}, 203 (2003).
\bibitem{exp} R.K. Lindell, J. Delahaye, M.A. Sillanp\"a\"a, T.T. Heikkil\"a,
E.B. Sonin, and P.J. Hakonen, Phys. Rev. Lett. {\bf 93}, 197002 (2004).
\bibitem{SN-T} E. B. Sonin, Phys. Rev. B {\bf 70}, 140506(R) (2004).
\bibitem{TN} J. Tobiska and Yu.V. Nazarov, Phys.Rev. Lett.  {\bf 93}, 106801
(2004).
\bibitem{Pek} J.P. Pekola, Phys.Rev. Lett.  {\bf 93}, 206601 (2004); J.P.
Pekola, T.E. Nieminen, M. Meschke, J.M. Kivioja, A.O. Niskanen, and J.J.
Vartiainen, {\sl ibid.} {\bf 95}, 197004 (2005).
\bibitem{grab} J. Ankerhold and H. Grabert, Phys.Rev. Lett.  {\bf 95}, 186601
(2005).
\bibitem{Scm} E.B. Sonin, cond-mat/0505424.
\bibitem{BF} V. Brosco, R. Fazio, F. W. J. Hekking,and J. P. Pekola, Phys. Rev. B {\bf
74}, 024524 (2006).
\bibitem{Aver} D.V. Averin, Yu.V. Nazarov, and A.A. Odintsov, Physica B {\bf 165\&166},
945 (1990).
\bibitem{SZ} G. Sch\"on and A.D. Zaikin, Phys. Rep. {\bf
198}, 237 (1990).
\bibitem{IN} G.L. Ingold and Yu.V. Nazarov, in {\sl Single Charge Tunneling,
Coulomb Blockade Phenomena in Nanostructures}, ed. by H. Grabert and M.
Devoret (Plenum, New York, 1992), p. 21.
\bibitem{Heik} T.T. Heikkil\"a, P. Virtanen, G. Johansson, and F.K. Wilhelm, Phys. Rev.
Lett. {\bf 93}, 247005 (2004).
\bibitem{TDPE} J. Hassel,  H. Sepp\"a,  J. Delahaye, and P. Hakonen, J.  Appl.  Phys.  {\bf 95}, 8059 (2004).
\bibitem{HPS} P.J. Hakonen,  A. Paila, and E.B. Sonin, cond-mat/0604479.
\bibitem{Naz} The relevance  of this approximation for the present analysis was pointed out by Yuli Nazarov. 
\bibitem{IG} G.L. Ingold and H. Grabert, Phys. Rev. Lett. {\bf 83}, 3721
(1999).
\end{thebibliography}
\end{document}